\documentclass[twocolumn,aps, prb, showpacs, preprintnumbers, amsmath, amssymb, a4paper, superscriptaddress]{revtex4-1}

\usepackage{graphicx}% Include figure files
\usepackage{dcolumn}% Align table columns on decimal point
\usepackage{bm}% bold math
\usepackage{color}
\preprint{APS/123-QED}

\newcolumntype{L}{>{$}l<{$}} % text mode "l" in an "array"

\begin{document}
	
\title{Electron trajectories and magnetotransport in nanopatterned graphene under commensurability conditions}

\author{Stephen R. Power}
\email{stephen.power@icn2.cat}
\affiliation{Center for Nanostructured Graphene (CNG), DTU Nanotech, Department of Micro- and Nanotechnology,
	Technical University of Denmark, DK-2800 Kongens Lyngby, Denmark}
\affiliation{Center for Nanostructured Graphene (CNG), Department of Physics and Nanotechnology, Aalborg University, Skjernvej 4A, DK-9220 Aalborg East, Denmark}
\affiliation{Catalan Institute of Nanoscience and Nanotechnology (ICN2), CSIC and The Barcelona Institute of Science and Technology, Campus UAB, Bellaterra, 08193 Barcelona (Cerdanyola del Vall\`es), Spain}
\affiliation{Universitat Aut\`onoma de Barcelona, 08193 Bellaterra (Cerdanyola del Vall\`es), Spain}

\author{Morten Rish{\o}j Thomsen}
\affiliation{Center for Nanostructured Graphene (CNG), Department of Physics and Nanotechnology, Aalborg University, Skjernvej 4A, DK-9220 Aalborg East, Denmark}

\author{Antti-Pekka Jauho}
\affiliation{Center for Nanostructured Graphene (CNG), DTU Nanotech, Department of Micro- and Nanotechnology,
	Technical University of Denmark, DK-2800 Kongens Lyngby, Denmark}

\author{Thomas Garm Pedersen}
\affiliation{Center for Nanostructured Graphene (CNG), Department of Physics and Nanotechnology, Aalborg University, Skjernvej 4A, DK-9220 Aalborg East, Denmark}
\email{tgp@nano.aau.dk}

\date{\today}% It is always \today, today,
%  but any date may be explicitly specified

%	\begin{tocentry}
%		
%			\includegraphics[width =1.0\textwidth]{toc}
%		
%	\end{tocentry}

	\begin{abstract}
		Commensurability oscillations in the magnetotransport of periodically patterned systems, emerging from the interplay of cyclotron orbit and the pattern periodicity, are a benchmark of mesoscopic physics in electron gas systems.
		Exploiting similar effects in 2D materials would allow exceptional control of electron behaviour, but is hindered by the requirement to maintain ballistic transport over large length scales.
		Recent experiments have overcome this obstacle and observed distinct magnetoresistance commensurability peaks for perforated graphene sheets (antidot lattices).
		Interpreting the exact mechanisms behind these peaks is of key importance, particularly in graphene where a range of regimes are accessible by varying the electron density. 
		In this work a fully atomistic, device-based simulation of magnetoresistance experiments allows us to analyse both the resistance peaks and the current flow at commensurability conditions.
		Magnetoresistance spectra are found in excellent agreement with experiment, but we show that a semi-classical analysis, in terms of simple \emph{skipping} or \emph{pinned} orbits,
		is insufficient to fully describe the corresponding electron trajectories.
		Instead, a generalised mechanism in terms of states bound to individual antidots, or to groups of antidots, is required.
		Commensurability features are shown to arise when scattering between such states is enhanced.
		The emergence and suppression of commensurability peaks is explored for different antidot sizes, magnetic field strengths and electron densities.
		The insights gained from our study will guide the design and optimization of future experiments with nanostructured graphene.
	\end{abstract}

\maketitle

\section{Introduction}
High-quality graphene samples with very large carrier mobilities provide an excellent platform to explore a range of physical phenomena emerging from the unique linear band structure near the Dirac point, such as Klein tunneling\cite{katsnelson2006chiral, young2009quantum}, the fractional quantum Hall effect\cite{bolotin2009observation} and electron lensing\cite{cheianov2007focusing, chen2016electron}.
The ability to continuously tune the electron density in graphene also allows for a range of mesoscopic phenomena, previously examined in semiconductor 2D electron gas systems, to be investigated more thoroughly and without the need for doping\cite{novoselov2004electric}.
For example, the study of quantised conductance in confined one-dimensional channels relies on controlling the ratio between the system width and Fermi wavelength.
Typically, this involves changing the width using gate-defined potentials at the channel edges, but can be achieved in graphene using a fixed-width nanoribbon and varying the Fermi wavelength by gating\cite{terres2016size}.
Ballistic graphene systems with variable electron densities are also ideal for probing mesoscopic classical physics effects -- the realization of transverse magnetic focusing is a key example of this and allows electrons to be guided between two leads by varying their cyclotron radius in a magnetic field\cite{taychatanapat2013electrically}.

\begin{figure}
	\centering
	\includegraphics[width =0.48\textwidth]{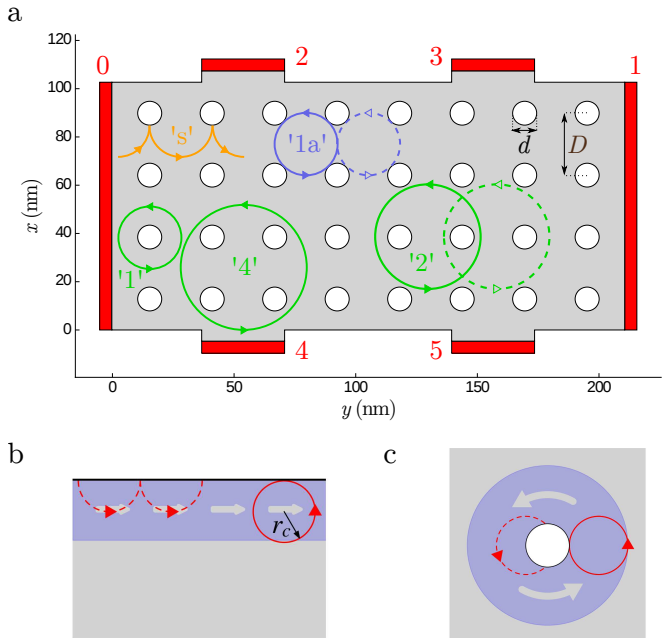}
	\caption{a) Simulated device geometry: the central graphene region (grey) contains a $8\times4$ antidot array with diameter $d\approx10$nm and separation $D\approx26$nm. 
		External leads ($0$--$5$) are shown in red. 
		Classical commensurability orbits are highlighted -- skipping (`s', orange), pinned around dots (`1', `2', `4', green) or pinned between dots (`1a', blue).
		b) Dispersive edge states in the quantum Hall regime occupy the blue shaded region near an edge. Classically, consecutive skipping orbits (dashed red trajectories) combine to give a net current flow  (grey arrows). The edge state region width, $2 r_c$, is given by the furthest extent of a cyclotron orbit which intersects the edge.
		c) A \emph{ring} of bound states, consisting of orbits trapped by the antidot, forms around an antidot in a magnetic field. These are non-dispersive unless coupled to other states by additional scattering.}
	\label{fig_geo}
\end{figure}

Antidot lattices provide perhaps the quintessential system for investigating two-dimensional mesoscopic phenomena\cite{weiss1991electron, Ando2000, PhysRevB.44.3447, PhysRevLett.68.1367, PhysRevB.55.16331, ando1999chaos, PhysRevB.51.7058, PhysRevB.53.7975, PhysRevB.53.7987, PhysRevB.56.4710, PhysRevB.56.15195}.
They are realised in electron gases by a periodic array of strong repulsive potentials which scatter electrons travelling through the system.
Hall bar transport measurements, performed in the presence of a perpendicular magnetic field, reveal a range of features.
When the cyclotron radius is large enough to scatter electrons between antidots, successive scatterings can connect both sides of the device, introducing backscattering and removing the quantised edge transport of the quantum Hall regime (QHR).
Most cyclotron radii in this regime lead to chaotic trajectories in the device bulk.
However, when the cyclotron radius is commensurate with important system length scales, prominent new peaks and (non-quantised) plateaux emerge in the longitudinal and Hall resistances, $R_{xx}$ and $R_{xy}$, respectively.
These are typically associated with semi-classical electron orbits which are \emph{pinned} around individual or groups of antidots (shown by green circles in Fig. \ref{fig_geo}a) or \emph{skipping} between nearby antidots (shown by the orange trajectory in Fig. \ref{fig_geo}a).
Fine Aharonov-Bohm (AB) oscillations on top of these commensurability features are associated with quantisation of the electron orbits\cite{PhysRevB.54.8021}.
For high enough magnetic fields, the cyclotron radius is too small to cause scattering between antidots, and the QHR is mostly restored apart from exponentially suppressed quantum tunneling between states localised near individual antidots\cite{Cserti_bound, mortenMGAL}.

Gate potentials are  inefficient at repelling electrons in graphene due to Klein tunneling\cite{katsnelson2006chiral}, and so graphene antidot lattices (GALs) instead consist of periodic perforations in graphene sheets.
The electronic properties of GALs and, in particiular, the dependence of band gaps on geometry has been the subject of intensive research\cite{Pedersen:GALscaling, ClarSextetGAL}. 
Moreover, the influence of magnetic fields on perfectly periodic GALs and isolated antidots has been studied at the level of full atomistic approaches\cite{pedersen2013hofstadter}, the Dirac approximation\cite{pedersen2012dirac}, or the simple gapped graphene model\cite{magnetooptical2011}. 
Experimental fabrication of GALs involves invasive techniques such as electron beam or block copolymer lithography\cite{Bieri2009, Kim2010, Bai2010, Kim2012, Shen2008, Eroms2009, Begliarbekov2011, Giesbers2012, Oberhuber2013, Xu2013}. 
A major issue is the deterioration of the graphene sheet quality and the difficulty in maintaining a uniform size and separation of antidots throughout the lattice.
Indeed, the band gap behaviour predicted for certain lattice geometries\cite{Pedersen:GALscaling, ClarSextetGAL, Ouyang2011, Liu2013, Vanevic2009, Tue:finiteGAL, Trolle-spinGAL, Pedersen2012a, GOALS, soren-mag} is particularly sensitive to small levels of geometric disorder which may not be possible to eliminate in experiment\cite{Yuan_GAL_disorder, Yuan_GAL_screening, HungNguyen2013, Ji2013, power2014electronic, fan2015electronic}.
Although such uniformity is not an essential ingredient for commensurability oscillations, invasive etching processes usually reduce the mean free path significantly so that electrons are principally scattered by defects and not antidots, thus suppressing commensurability effects.
By encapsulating graphene in hexagonal boron nitride (hBN) during the nanopatterning step, Sandner \textit{et al}.\cite{sandner2015ballistic} demonstrate that the sample quality can be protected and consequently they observed pronounced commensurability peaks which they associate with orbits around 1, 2, or 4 antidots in a square lattice.
A contemporaneous study by Yagi \textit{et al}.\cite{yagi2015ballistic}, using a different etching technique, found commensurability features in triangular lattices with smaller mean free paths, suggesting that scattering between nearest neighbour antidots plays a key role. 

To clarify the exact mechanisms behind these commensurability peaks and determine the electron trajectories through such systems, we perform large-scale atomistic transport simulations of graphene antidot lattice devices.
An excellent agreement is found, both in the relative positions and magnitudes of commensurability peaks, with recent experimental results \cite{sandner2015ballistic}.
Furthermore, by mapping the current flow we can identify the electron trajectories associated with each peak.
The classical picture of skipping or pinned orbits is too simple to fully describe the resulting electron flow patterns.
An alternative analysis, in terms of scattering between states bound to individual antidots, shows that a generalised skipping orbit picture can explain the first two commensurability peaks. 
Higher order peaks are understood in terms of quasi-pinned orbits around groups of antidots.
Both mechanisms act to divert electrons away from the sample edges and into the bulk, and so remove the ballistic edge transport of the quantum Hall regime.
Finally, we examine the emergence and suppression of the peaks at different antidot sizes and the experimentally relevant transition between classical and quantum regimes.

\section{Methods}
\label{sec:methods}
We consider a six probe Hall bar structure as shown schematically in Fig. \ref{fig_geo}a.
The main device region is constructed from a $\sim100$nm wide zigzag nanoribbon with a $960$-atom unit cell.
The six external leads consist of semi-infinite nanoribbons.
A few atomic rows from the top and bottom armchair nanoribbon probes are included in the device region for mapping purposes.
The main device region consists of a $4\times8$ array of antidots with a center-to-center separation of $D\approx26$nm.
We focus on antidots with diameter $d\approx10$nm, as shown in this schematic, but also consider antidots of different sizes. 
The total number of atoms in a typical simulation is between 750,000 and 950,000.

The electronic structure of graphene is described by a single $\pi$-orbital tight-binding Hamiltonian
$
H = \sum_{<ij>} t_{ij} (B) \, {\hat c}_{i}^\dag \, {\hat c}_{j}\;\,,
%\label{hamiltonian}
$
where the sum is taken over nearest-neighbour sites only.
The nearest-neighbouring hopping parameter $ṭ_{ij} (B)$ takes the value $ṭ_{ij} (0) = t \equiv -2.7\mathrm{eV}$ in the absence of a magnetic field.
Throughout this work we will use $|t|$ as the unit of energy.
The effect of a magnetic field is included using the Peierls' phase approach, which places a field-dependent phase factor into the hopping parameters
\begin{equation}
t_{ij} (\mathbf{B}) = t_{ij} (0) e ^{\frac{2 \pi i e}{h} \Theta_{ij}},
\label{peierl-hoppings}
\end{equation}
where
%\begin{equation}
$\Theta_{ij}   = \int_{\mathbf{r}_i}^{\mathbf{r}_j} \mathbf{A}(\mathbf{r}^\prime) \cdot \mathrm{d}
\mathbf{r}^\prime$
%\label{peierls-phase}
%\end{equation}
and $\mathbf{r}_i$ is the position vector of site $i$ and we have a choice of gauge fields $\mathbf{A}$ which give $ \mathbf{B} = \nabla \times \mathbf{A} =B\hat{z}$.
%Generally
The commonly used Landau gauge $\mathbf{A}_0 = - B y \hat{x}$ maintains periodicity along one direction.
However, a more complicated prescription, taking advantage of periodicity in the graphene leads in both $x$ and $y$ directions\cite{barangerstone}, is employed for our Hall bar calculations and is outlined in more detail in Appendix \ref{appendix-phases}.
Removing Carbon atoms from the antidot regions corresponds to the removal of the associated rows and columns from the system Hamiltonian.
We also remove dangling atoms with only a single remaining neighbouring atom.
Any dangling $\sigma$ bonds for a Carbon atom with only two neighbouring Carbon atoms are assumed to be passivated by Hydrogen so that the $\pi$ bands are unaffected.

The zero-temperature currents $I_p$ and potentials $V_p$ in each lead $p$ are related via the multi-terminal Landauer-Buttiker relation
$
I_p = \frac{2e}{h} \sum \left( T_{qp} V_p - T_{pq}V_q\right),
$
where the transmissions are given by the Caroli formula \cite{caroli1971direct}
$
T_{pq}  = \mathrm{Tr}  \left[ \,G^R  \, \Gamma_q  \, G^A \, \Gamma_p  \,\right] \,.
$
Here, $G^R$ and $G^A$ are the retarded and advanced Green's functions, respectively, and $\Gamma_p$ is the broadening matrix associated with lead $p$.
Defining $\tilde{T}_{pq} = \delta_{pq} \sum_r T_{rp} - T_{pq}$ gives a direct relation  $\mathbf{I} = \mathbf{\tilde{T}} \mathbf{V}$, where $\mathbf{I}$ and $\mathbf{V}$ are column-vectors.
To calculate the longitudinal and Hall resistances, we fix a potential difference between the left and right leads, and set the net current in the top and bottom floating probes to zero.
Solving for the net current flowing through the system, $I$, and the potentials at the top and bottom probes $V_{2-5}$ yields the longitudinal and Hall resistances 
\begin{equation}
R_{xx}  = \frac{V_2 - V_3}{I}\,, \qquad\qquad  R_{xy}  = \frac{V_2 - V_4}{I} \,.
\end{equation}
The Green's functions required are calculated using efficient recursive techniques to return the matrix elements required\cite{Lewenkopf2013, patchedGF}.

Fine oscillations, similar to those noted in Ref. \cite{sandner2015ballistic}, are observed superimposed on the simulated commensurability peaks.
A Fourier analysis confirms these to be mostly Aharanov-Bohm in origin, with a high weight at periods  $\Delta B \sim \frac{h}{e A}$, where $A$ is the area of the periodic antidot unit cell.
To emphasise the commensurability peaks, which are the key focus of this work, a small amount of geometric disorder (random fluctuations of $\lesssim 1$nm  in the positions and radii of individual antidots) is introduced and an average taken over five such instances.
Individual and averaged configurations are shown later in Figs.  \ref{fig_sizedep} and \ref{fig:efdep}.

To produce spatial maps of the current density at particular values of magnetic field and Fermi energies we calculate the bond currents\cite{cresti-currents, nikolic-currents, Lewenkopf2013} between neighbouring sites $i$ and $j$ under the Hall bar constraints above and find
\begin{equation}
I^{(n-eq.)}_{ij} \sim -\sum_{p\ne 1} (V_p - V_1) \mathrm{Im} \, \left( t_{ij}(B) (G^R \Gamma^p G^A)_{ji} \right) \,.
\label{eqn:final_bond_currents2}
\end{equation}
Further details are given in Appendix \ref{appendix-currents}.
Current heat maps in this work show the spatial distribution of current in the system, and represent the magnitude of the local current using brighter (darker) colours for larger (smaller) magnitudes.
Arrows are superimposed in some zoomed current maps to illustrate the current flow direction.
In both cases a spatial averaging is applied to produce current map plots with suitable resolution and clarity.

\section{Results and discussion}

\begin{figure*}
	\centering
	\includegraphics[width =0.94\textwidth]{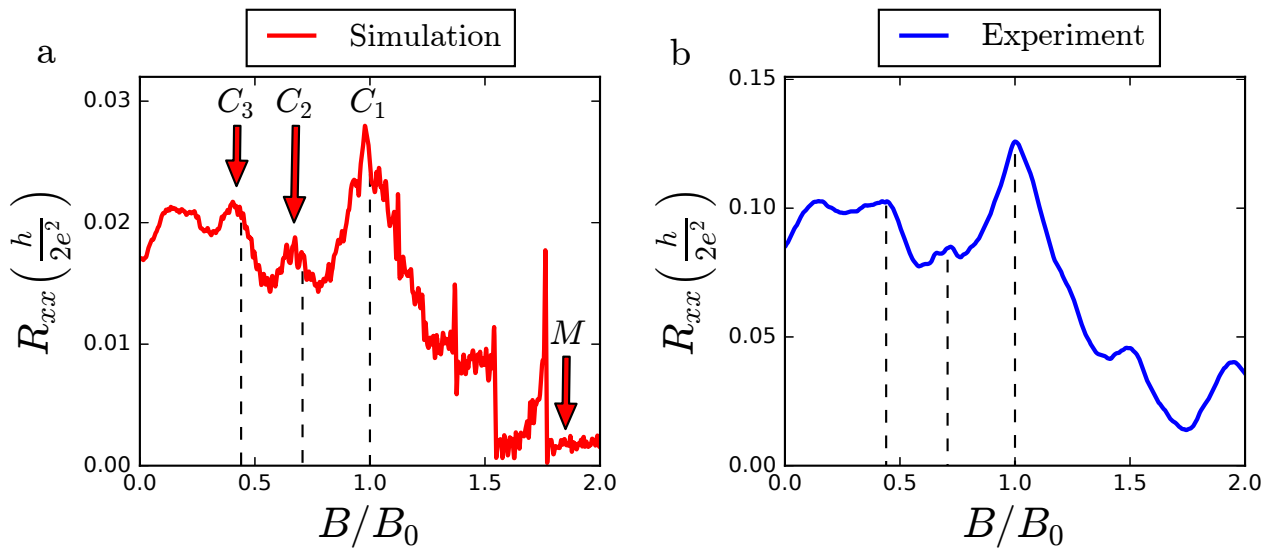}
	\caption{ a) $R_{xx}$  from a Hall bar simulation for the system in Fig. \ref{fig_geo}a at  $\lambda_F\approx 3$ nm $<D,d$, averaged over five instances with mild geometric disorder. The three commensurability peaks ($C_{1,2,3}$) and modified quantum Hall regime (M) are shown by the red arrows. b) The experimental $R_{xx}$ result reproduced from Sandner \textit{et al}. \cite{sandner2015ballistic} -- an excellent match is noted with the simulation.}
	\label{fig_rxx}
\end{figure*}

\subsection{Magnetoresistance commensurability peaks}
\label{sec:peaks}
The six probe Hall bar device employed in our simulations is shown in Fig. \ref{fig_geo}a, together with schematics of different semi-classical pinned and skipping orbits. 
We consider a system with antidot separation $D\approx 26$nm and diameter $d\approx 10$nm, with mild geometric disorder in antidot size and position.
The simulated longitudinal resistance $R_{xx}$ of this systems is shown in Fig. \ref{fig_rxx}a as a function of the perpendicular magnetic field $B$.
The Fermi wavelength is set much smaller than GAL length scales ($\lambda_F = \frac{h v_F}{E_F} \sim3$nm for  $E_F=0.4|t|$) to ensure we are outside the quantum regime, i.e. $\lambda_F\simeq D,d$.
The cyclotron radius, determined by the magnetic field strength, is given by
\begin{equation}
r_c = \frac{E_F}{v_F e B} \,.
\label{cyclo_rad}
\end{equation}
The fully atomistic nature of our calculations limits the size and separation of our antidots, hence quite a large $E_F$ is used to achieve an appropriate Fermi wavelength.
A large magnetic field is then required in Eq. \ref{cyclo_rad} to enter the commensurability regime $r_c \sim D$.
In fact, the condition $r_c = \frac{D}{2}$, associated with the primary commensurability peak occurs at $B_0 \approx 97$T for our system.
Experimental systems typically have $D \gtrsim 100$nm and $d \sim 30$nm, so that the commensurability regime is accessible for smaller $E_F$ and with $B<10$T.
The experimental and theoretical systems are directly related by scaling -- from Eq. (\ref{cyclo_rad}) we expect $B_0 \sim \frac{1}{D^2}$ when the ratio $\frac{\lambda_F}{D}$ is kept fixed -- and the same qualitative processes determine electron behaviour in each. 
Indeed, the experimental peak observed in Ref. \cite{sandner2015ballistic} at $B_0\approx3.6$T and $\lambda_F\approx21$nm corresponds to $r_c\approx 54$nm -- very near the expected value $r_c = 50$nm.
Magnetic fields are plotted here in terms of $B_0$ for easy comparison of results at different scales and energies.
In our simulations $B_0$ is calculated using Eq. \ref{cyclo_rad}.

The $R_{xx}$ curve in Fig. \ref{fig_rxx} contains (at least) three prominent peak features, which we denote as commensurability peaks $C_1$, $C_2$ and $C_3$.
$C_1$ occurs almost precisely at the expected $B_0$.
The validity of our results is confirmed by a direct comparison with the experimental data of Sandner \textit{et al}\cite{sandner2015ballistic} in a system with $D\sim100$nm, $d\sim30$nm, which is reproduced in Fig. \ref{fig_rxx}b.
$B_0$ is set as the peak maximum for the experimental data.
An excellent agreement is noted between simulation and experiment, suggesting a universal behaviour at different scales once the ratios between cyclotron radii and system dimensions are similar.
The three peaks in $R_{xx}$ coincide with new step-like features in $R_{xy}$ (see Appendix \ref{appendix-rxy}), a feature that is also visible in the experiment  and in previous experimental and theoretical studies in 2DEGs\cite{weiss1991electron, Ando2000}. 
Shubnikov-de Haas oscillations are visible for high fields in both curves. 
Their spacing follows the Landau levels, and not the pattern geometry, and so differs between the experimental and simulated systems. 

In an unpatterned sample, $R_{xx}$ displays quantum Hall behaviour and is nonzero only in two scenarios: i) at discrete magnetic field values corresponding to bulk Landau levels (LL), and ii) at small magnetic field values where edge states are not tightly enough confined.
In both cases, a breakdown of edge-only ballistic transport occurs due to the presence of bulk states allowing transmission between different edges.
A similar mechanism lies behind the enhancement of $R_{xx}$ in  antidot devices.
In the absence of edges or other scatterers, electrons in a magnetic field are localised.
This is viewed semi-classically in terms of closed orbits with cyclotron radius $r_c$, whereas quantum-mechanically electrons occupy discrete LLs whose spatial extent is determined by the magnetic length $l_B$.
In the presence of an edge, the confining potential gives dispersion to nearby states so that conducting edge channels are formed for each LL.
These states allow propagation in one direction along each edge, with the total transmission given by the number of occupied LLs.
Semi-classically, all orbits whose centers lie within $r_c$ of the edge should scatter from it and form `skipping orbits' as shown by the dashed red trajectory in Fig. \ref{fig_geo}b.
In fact, the total width of the conducting edge region is excellently approximated by $2 r_c$ -- the furthest possible reach of an orbit which intersects the edge.
This edge state region is shaded blue in Fig. \ref{fig_geo}b, with the net current flow shown by the grey arrows.
Dispersive edge state formation is thus associated with hybridisation or coupling between multiple neighbouring localised states due to scattering.

In a similar manner, the edge of an antidot also couples neighbouring localised states to create a \emph{ring} of radius $2r_c$ beyond the edge of the antidot (i.e. total radius $2r_c + \frac{d}{2}$ from the antidot center).
This region (shaded blue in Fig. \ref{fig_geo}c) encloses all the semi-classical orbits which intersect an antidot.
The net circulation direction is the same as that of single localised orbits and is shown in Fig. \ref{fig_geo}c by grey arrows.
We note that these regions are not the same as the `1'-type pinned orbit (Fig. \ref{fig_geo}) which consist of single semi-classical orbits with radius $r_c$ encircling, but not scattering from, an antidot.
Unlike QHR edge states, antidot ring states are non-dispersive and are bound to individual antidots in the absence of additional scattering centres.
We now consider the current flow in antidot Hall bar devices as the cyclotron radius increases (magnetic field \emph{decreases}).

%\begin{figure}
%	\centering	
%	\textbf{MQHR}
%	
%	\vspace{0.02\textwidth}
%	
%%	\hspace{0.37\textwidth}
%%	\textbf{$C_1$}
%	\includegraphics[width =0.47\textwidth]{fig_parts/curr1}
%	\caption{Schematics and current maps for the modified quantum Hall (MQHR)  
%		a, d) Schematics showing extent of edge and/or ring states (blue shaded areas) in the given regime, net current flow directions (grey, pink arrows), and individual semi-classical trajectories (red dashed and solid lines). The thicker green circle in d) shows the expected size of the `1a' type pinned orbit at this magnetic field value.
%		b, e) Current heat maps of the entire device.
%		c, f) Zoomed region of heat maps in (b,e) with current flow directions illustrated by arrows
%	}
%	\label{fig:currents}
%\end{figure}

\subsection{Current flow for large magnetic fields}

\begin{figure*}
	\centering	
	\textbf{MQHR}
	\hspace{0.37\textwidth}
	\textbf{$C_1$}
	\includegraphics[width =0.98\textwidth]{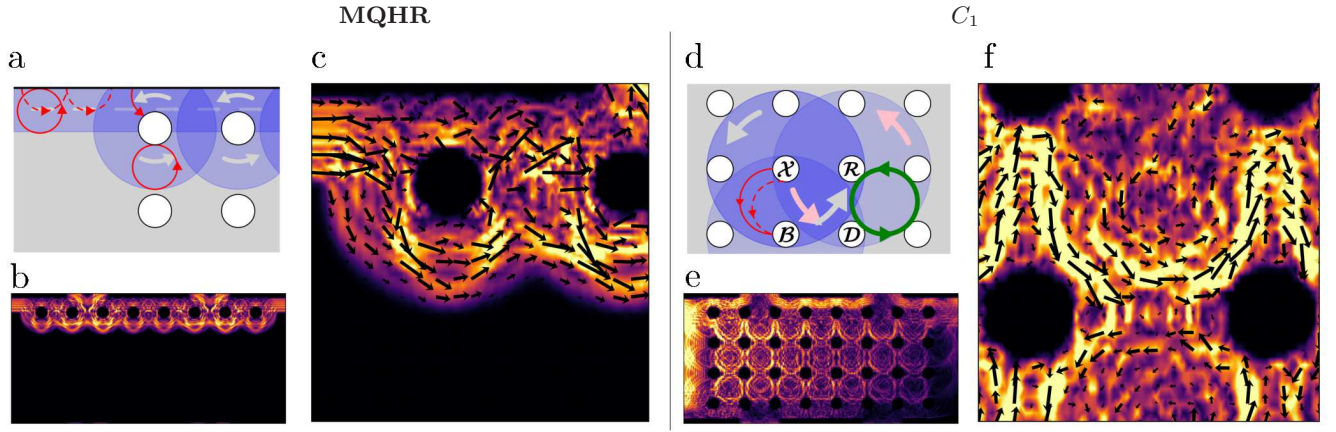}
	\caption{Schematics and current maps for the modified quantum Hall (MQHR)  and first commensurability peak ($C_1$) regimes.
		a, d) Schematics showing extent of edge and/or ring states (blue shaded areas) in the given regime, net current flow directions (grey, pink arrows), and individual semi-classical trajectories (red dashed and solid lines). The thicker green circle in d) shows the expected size of the `1a' type pinned orbit at this magnetic field value.
		b, e) Current heat maps of the entire device.
		c, f) Zoomed region of heat maps in (b,e) with current flow directions illustrated by arrows
	}
	\label{fig:currents}
\end{figure*}

There is no interaction between the QHR edge states and the antidot region for very large fields, and the current flow is unchanged relative to an unpatterned device. 
Changes in bulk state distribution, due to broken LL degeneracy, can manifest as a broadening of Shubnikov-de Haas oscillations in $R_{xx}$.
Significant changes from the QHR occur when $r_c$ is increased so that the ring associated with an antidot intersects another antidot or the device edge.

%The separation between the top row of antidots and the system edge in our geometry is smaller than that between neighbouring antidots.
As $B$ is decreased, $r_c$ becomes large enough to allow hybridisation between quantum Hall edge states and the rings associated with the first row of antidots.
This modified quantum Hall regime (MQHR) is considered on the left of Fig. \ref{fig:currents}, where $B\approx1.85 B_0$.
At this field, the rings associated with the top row of antidots intersect the top edge, and similarly the edge state region intersects these antidots (Fig. \ref{fig:currents}a).
However, while the rings of neighbouring antidots overlap, they do not intersect the neighbouring antidots themselves.
The current map in Fig. \ref{fig:currents}b, and the zoom in Fig. \ref{fig:currents}c, show that the incoming edge current from the left is split at the first antidot it encounters.
The current then flows in a broader edge channel encompassing the first row of antidots, and which is a superposition of the QHR edge state region and the rings associated with top row antidots.
New dispersive states, which now also flow around the first row of antidots, are formed by coupling between states in these regions due to edge and antidot scattering.
The key factor is that the overlap region between the QHR edge state and an antidot ring contains scatterers, namely the edge and antidot themselves.
We also note that the overlap between neighbouring rings on the top row is intersected by the top edge, so that states previously bound to individual antidots can couple to each other due to the edge.
No such hybridisation occurs between the bound states associated with antidots in lower rows, as their overlap regions are devoid of scatterers, so these states remain localised and do not contribute to current flow in the system.

\subsection{Current flow near the principal peak}
At larger $r_c$, antidot rings impinge not only on each other, but on the edges of the neighbouring antidots themselves.
Semiclassically, this begins once a single orbit can cross the neck between two neighbouring antidots, \emph{i.e.}, from $2r_c = D - d$, corresponding to $B\approx1.6B_0$ for the aspect ratio here.
$C_1$ is prominent in our simulations between $0.8 B_0 < B < 1.2 B_0$, with a particularly sharp peak at $0.96 B_0< B<1.01 B_0$, suggesting that the migration of electrons between neighbouring rings contributes significantly for $B\lesssim1.2B_0$.
The extent of an antidot ring region at $B=B_0$ is shown by the dark blue shaded region around antidot $\mathcal{X}$ in Fig. \ref{fig:currents}d.
The grey arrows show the net current direction in this ring.
The lighter shaded regions show the rings of the antidots below ($\mathcal{B}$) and to the right ($\mathcal{R}$) of $\mathcal{X}$, and pale pink arrows show the current direction around $\mathcal{R}$.
The commensurability condition $2 r_c = D$, associated with simple skipping orbits (solid red curve), is also the condition for an antidot ring to entirely enclose the antidots nearest to it. 

%\begin{figure}
%	\centering	
%	\textbf{$C_1$}
%	
%	\vspace{0.015\textwidth}
%	
%	%	\hspace{0.37\textwidth}
%	%	\textbf{$C_1$}
%	\includegraphics[width =0.47\textwidth]{fig_parts/curr2}
%	\caption{As Fig. \ref{fig:currents}, but for $B$ corresponding to the principal commensurability peak $C_1$.
%	}
%	\label{fig:currents2}
%\end{figure}

\begin{figure*}
	\centering	
	\textbf{$C_2$}
	\hspace{0.45\textwidth}
	\textbf{$C_3$}	
	\includegraphics[width =0.98\textwidth]{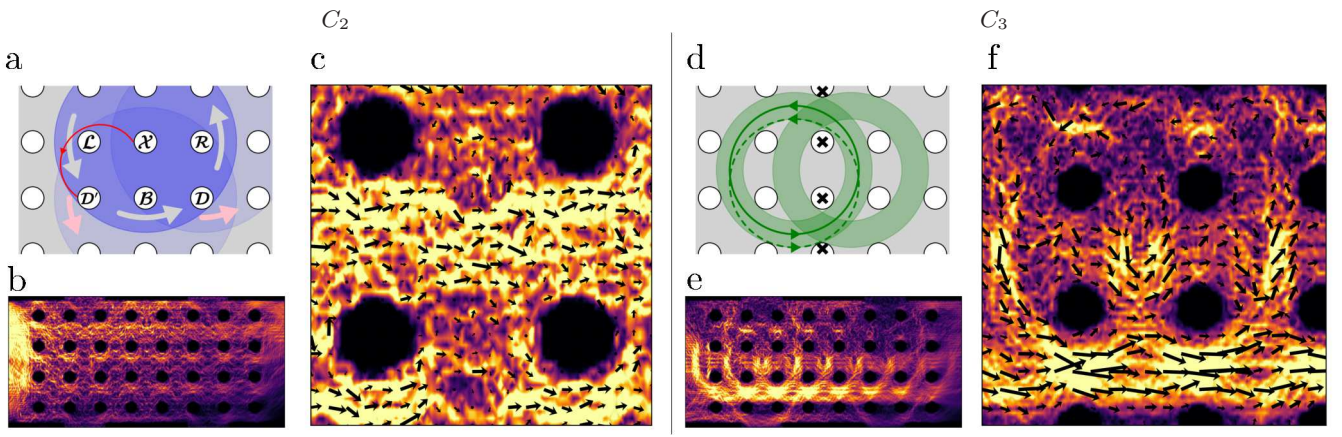}
	\caption{As Fig. \ref{fig:currents}, but for $B$ corresponding to commensurability peaks $C_2$ (left) and $C_3$ (right).
		a, d) Schematics showing extent of ring states (blue shading) and current flow directions (grey, pink arrows), and individual classical trajectories (red and green lines). The shaded green area in d) shows the range of `4' type quasi-pinned orbits (green, dashed line) which intersect one of the surrounding antidots .
		b, c, e, f) Current heat maps and zooms
	}
	\label{fig:currents2}
\end{figure*}

Let us examine the coupling between states associated with neighbouring antidots as $r_c$ increases.
In the range $ D- d <2 r_c < D- \frac{d}{2}$ (or $1.6B_0 \gtrsim B \gtrsim 1.2 B_0$), the ring around $\mathcal{X}$ develops from tangentially touching $\mathcal{B}$ to encompassing its entire top half.
The portion of the antidot within the ring acts as a scatterer which can couple states associated with $\mathcal{X}$ and $\mathcal{B}$.
However, near the top half of $\mathcal{B}$, states within their respective rings have opposite orbital directions, e.g. states associated with $\mathcal{X}$ are on average right-moving, whereas those of $\mathcal{B}$ are left-moving.
This makes it harder to scatter between rings, leading to preferential scattering between states within the same ring and only a weak coupling between states from neighbouring antidots. 
Current maps for $B$ and $r_c$ in this range show that most of the current is carried by states near the top edge and top row of antidots, similar to Fig. \ref{fig:currents}b, c, with only a minor contribution from lower rows.
The sharp $C_1$ peak emerges for $ D- \frac{d}{2} \lesssim 2 r_c < D$ (or $1.2B_0 \gtrsim B \gtrsim 1.0 B_0$), when the ring of $\mathcal{X}$ intersects the lower half of $\mathcal{B}$.
Here states in both rings are predominantly right-moving and a much stronger coupling is established.
In semi-classical terms, the high-field edge of $C_1$ corresponds to the smallest orbits which can intersect both $\mathcal{X}$ and the lower half of $\mathcal{B}$ (the red dashed curve in Fig. \ref{fig:currents}d).
$C_1$ reaches a maximum at $B_0$, when the ring of $\mathcal{X}$ first completely surrounds antidot $\mathcal{B}$, due to two factors: i) the relative scattering area of $\mathcal{B}$ in the ring overlap region reaches a maximum near here (there is a small correction towards the high field side that varies with aspect ratio), ii) the new scattering regions introduced near $B=B_0$ are particularly good at coupling states in neighbouring rings due to aligned flow directions.
As $r_c$ increases further, the relative scattering area in the overlap region decreases rapidly, together with the relative measure of new hybridised states.
Semi-classically, although we introduce additional orbits from $\mathcal{X}$ to $\mathcal{B}$, we introduce others which avoid $\mathcal{B}$ altogether.

Examining the current flow patterns at $C_1$ in Fig. \ref{fig:currents}e, f, it is tempting to interpret those between the top two rows of antidots as `s'-type skipping orbits bouncing from the undersides of antidots in the top row.
In the lower rows, an interpretation in terms of overlapping `1a'-type pinned orbitals or superimposed horizontal and vertical `s' skips seems plausible.
However, the ``skipping''-type trajectory is more accurately interpreted as a hybridisation between neighbouring rings, such as those of $\mathcal{X}$ and $\mathcal{R}$ in Fig. \ref{fig:currents}d.
The resultant current path follows the pink and grey arrows, and resembles a simple skipping orbit.
The circular flow pattern in lower rows emerges from a superposition of this feature with an inverted flow emerging from the row of antidots underneath.
This analysis is supported by examining smaller antidot diameters, discussed below, where the net flow pattern takes on the triangular shape suggested by the pink and grey arrows in Fig. \ref{fig:currents}d.
This shape follows that of the overlap region between neighbouring rings.
The circular appearance of trajectories in Fig. \ref{fig:currents}e, f emerges from two possible factors.
Firstly, the aspect ratio here is $\frac{d}{D}=0.385$ -- very near to the critical value $\frac{d}{D}=\sqrt{2}-1$ where a ring begins to intercept second nearest neighbour (2NN) antidots.
This is evident in Fig. \ref{fig:currents}d, where the ring around $\mathcal{X}$ approaches 2NN antidots (those with a diagonal separation, including $\mathcal{D}$).
Additional scattering alters the current path, and can also deflect electrons between left-to-right and top-to-bottom commensurabilty channels.
Secondly, at these aspect ratios it is no longer possible to have completely pinned orbits of type `1a' between antidots.
As shown by the green orbit in Fig. \ref{fig:currents}d, these orbits now impinge upon their surrounding antidots and join the associated rings.
Current flow at this aspect ratio therefore has a more circular character due to the hybridisation of `1a'-type orbits with the conducting channels.

\subsection{Current flow for higher order peaks}
The position of $C_2$ at $0.6B_0 < B < 0.75B_0$, with a maximum at $B\approx 0.68 B_0$, suggests an extension of the $C_1$ mechanism to 2NN antidots.
At $2 r_c = \sqrt{2}D$ ($B\sim 0.71B_0$) an antidot ring entirely encompasses 2NN antidots (Fig. \ref{fig:currents2}a).
This gives direct coupling between states bound to antidots separated diagonally, but it is also the condition for maximum indirect coupling between \emph{nearest neighbouring} antidots due to scattering from a third antidot.
In Fig. \ref{fig:currents2}a, this corresponds to increased coupling between right-moving states bound to $\mathcal{X}$ and its neighbour to the right, $\mathcal{R}$, due to the presence of $\mathcal{D}$ (and also $\mathcal{B}$).
Similarly, coupling is increased between down-moving states around $\mathcal{X}$ and $\mathcal{B}$ (due to $\mathcal{D^\prime}$ and $\mathcal{L}$).
The grey (around $\mathcal{X}$) and pink (around $\mathcal{R}$ and $\mathcal{B}$) arrows suggest the formation of horizontal and vertical conducting channels between antidots.
Clear signatures of horizontal channels are observed in Fig. \ref{fig:currents2}b, c.
The high-field onset of $C_2$ at $B\sim 0.75B_0$ is below that expected by assuming  significant enhancement occurs once orbit sizes allow scattering from the preferential side of the antidot.
However, unlike the $C_1$ case, other antidots block many of these trajectories, particularly for smaller orbits, resulting in a narrower $C_2$ peak.
2NN antidots thus influence commensurability peaks in two ways: first by enhancing certain $C_1$ channels at small $r_c$, and secondly by giving rise to $C_2$ channels at higher $r_c$.

At even larger $r_c$, it is difficult to associate rings with individual antidots.
Each ring surrounds several antidots and has large overlaps with many other rings, leading to a large degree of delocalisation.
This is consistent with a wide distribution of dispersive bulk states, and a reduced contribution from ballistic edge transport to transmission between probes 2 and 3.
$R_{xx}$ therefore remains reasonably large and constant over the low field range.
This prediction is confirmed in both simulation and experiment, further validating our analysis in terms of antidot quasi-bound states.
$C_3$ occurs as a reasonably wide peak superimposed on this low-field plateau between $0.3B_0 \lesssim B \lesssim 0.5B_0$, corresponding to $3.33D \gtrsim 2 r_c \gtrsim 2D$.
`4'-type pinned orbits, like that shown in solid green in Fig. \ref{fig:currents2}d,  are predicted occur in a range centered around $B\sim0.44B_0$.
This orbit does not intersect an antidot, so is not expected to couple with other states to become dispersive.
However, a range of `quasi-pinned' orbits with the same radius (such as that shown in dashed green), but asymmetrically positioned relative to the antidots they surround, do intersect antidots.
These orbits collectively fill the green shaded areas to form a wide loop of hybridised states, in addition to individual localised orbits, circulating in this region.
Furthermore, the antidots marked $\mathbf{x}$ hybridise states in neighbouring four-antidot loops.
Electrons can therefore migrate through the device by circling consecutive 4-antidot loops in turn.
Such behaviour is clearly visible in Fig. \ref{fig:currents2}e, f, where a migrating current around groups of four antidots in the central rows of the device is noted.
The net effect is to carry current away from the edges and into the device bulk, increasing the resistance between pairs of edge probes.
At nearby fields, clear four-ring orbits are less visible, but the net effect of increased current in the bulk at the expense of edge channels remains evident.

%\begin{figure*}
%	\centering
%	
%	\begin{tabular}{ m{0.4\linewidth}  m{0.6\linewidth}}
%		a  & b  \\
%		\hspace{0.04\textwidth}
%		\includegraphics[width =0.28\textwidth]{fig5a_current} 		 &
%		\hspace{0.03\textwidth} \includegraphics[width =0.5\textwidth]{fig5b_sizes}
%	\end{tabular}
%	
%	\caption{a) Current map zoom at $C_1$ for $d\approx7nm$ showing a current  path with a more angular profile than that in Fig. \ref{fig:currents}f, corresponding to flow between rings of neighbouring antidots.
%		b) $R_{xx}$ for a range of antidot diameters, each dark curve in an average over individual instances of mild geometric disorder (light curves). Vertical lines show the expected peak positions.}
%	\label{fig_sizedep}
%\end{figure*}

\subsection{Commensurability peak mechanisms}
\label{sec:disc}
\begin{figure}
	\centering
	\includegraphics[width =0.5\textwidth]{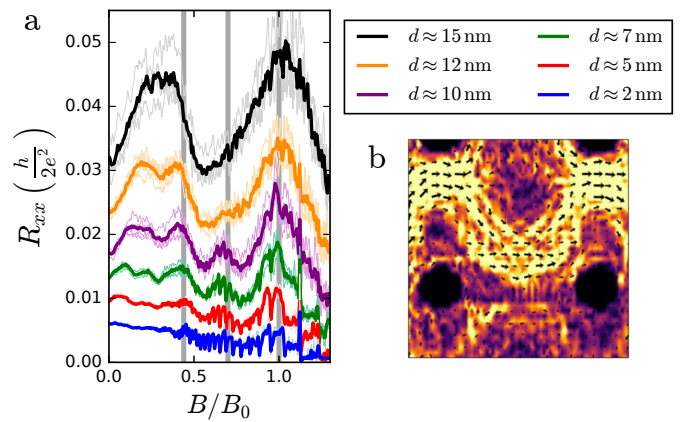} 		
	\caption {a) $R_{xx}$ for a range of antidot diameters, each dark curve in an average over individual instances of mild geometric disorder (light curves). Vertical lines show the expected peak positions.
b) Current map zoom at $C_1$ for $d\approx7nm$ showing a current  path with a more angular profile than that in Fig. \ref{fig:currents}f, corresponding to flow between rings of neighbouring antidots.}
	\label{fig_sizedep}
\end{figure}

%\subsection{General note on commensurability peak mechanisms}
The mechanisms behind $C_1$ and $C_2$ can both be considered generalisations of the semi-classical `skipping' orbit picture.
In both cases, current percolates through a network of quasi-localised states surrounding individual antidots.
At the principal peak $C_1$, coupling between states at neighbouring antidots is enhanced due to direct scattering by one antidot of electrons associated with the other.
The net trajectories resemble semi-classical runaway orbits, but current profiles for small antidots have an angular shape, as shown in Fig. \ref{fig_sizedep}b.
This is consistent with hybridisation between the rings localised around neighbouring antidots and not with a simple single skip mechanism.

In the case of $C_2$, the peak is not solely due to a runaway orbit scenario between pairs of second nearest neighbours, but is also associated with runaway orbits between neighbouring antidots, due to additional scattering from 2NN antidots.
We can rule out dominant contributions from `2'-type quasi-pinned orbits as the expected peak position for such orbits ($B\sim0.62B_0$) is to the low-field side of the peak in our simulations.

Meanwhile, the $C_3$ peak emerges due to quasi-pinned type orbits, whose net effect is to divert electrons away from edge states and into bulk channels.
Simulation of higher order orbits, similar to $C_3$ but pinned around larger groups of antidots, is non-trivial in our geometry as orbits around 9 or more antidots begin to intersect the edges of the device.
Such orbits are also difficult to achieve experimentally as even greater mean-free paths are required.
Nonetheless, enhancement of $R_{xx}$ should in principle occur at these radii also. 
Indeed, a slight peak is present at $\sim 0.1 B_0$ in both theory and experiment.

\begin{figure*}
	\centering
	
%	\begin{tabular}{m{0.61\linewidth} m{0.30\linewidth}}
%		\includegraphics[width =0.61\textwidth]{fig6ad_efs} &
%		e \includegraphics[width =0.30\textwidth]{fig6e_currents}
%	\end{tabular}
	\includegraphics[width =0.92\textwidth]{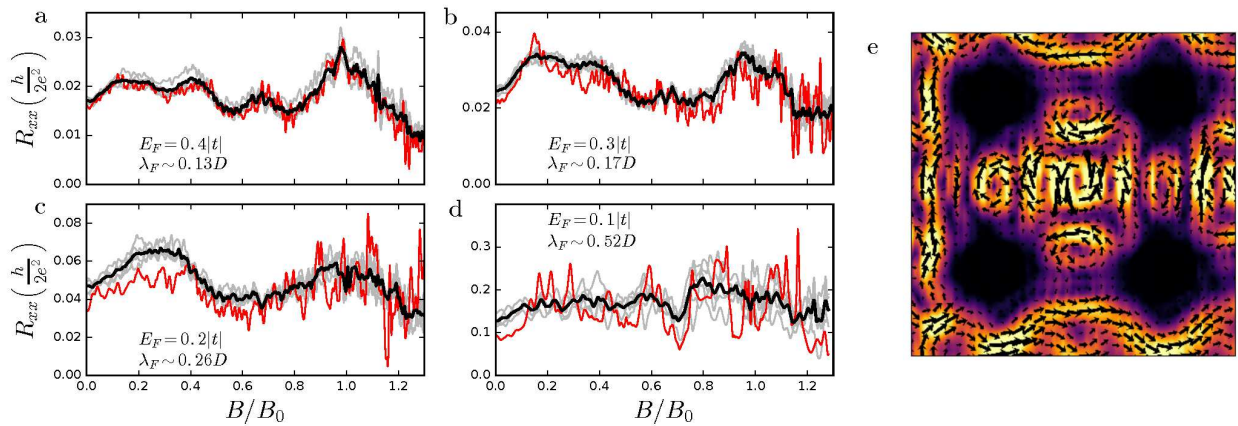} 
	\caption{a-d) $R_{xx}$ at 4 different Fermi energies, corresponding to an increase in Fermi wavelength. Red curves show the pristine system, whereas black curves are the result of averaging 5 configurations with mild position and radial disorder, with individual configurations in grey.
		e) Current map at $B=B_0$ for $E_F=0.1|t|$. Skipping orbits are notably absent, replaced by vortices induced by quantum interference. }
	\label{fig:efdep}
\end{figure*}
%\subsection{Geometry effects}
\subsection{Geometry and gating effects}
In Fig \ref{fig_sizedep}a, the antidot diameter $d$ is varied between $5$nm and $14$nm, equivalent to aspect ratios $\frac{d}{D}\sim0.2$ to $0.58$.
Each bold curve is an average over several instances of mild geometric disorder, with individual configurations shown by the lighter curves.
These curves are not shifted relative to each other along the $R_{xx}$ axis, but rather the magnitude of $R_{xx}$ increases significantly with antidot size.
Larger antidots act as stronger scatterers and divert more incoming edge current into the bulk of the device.
$C_1$ is prominent at $B=B_0$ for all but the smallest antidot diameters considered, but is particularly sharp near the $d\approx10$nm case considered in detail in the previous section.
This may be due to the enhanced scattering caused by 2NN antidots at this aspect ratio.
We note that this antidot size also gives the clearest $C_2$ peak.
At smaller sizes, the peak is somewhat difficult to detect due to the superposition of quantum oscillations, which are related to the underlying Shubnikov-de Haas oscillations\cite{PhysRevB.56.15195}.
At larger sizes, it is less prominent due to the blocking effect of nearest neighbour antidots and is largely swallowed by the broadening $C_1$ peak.
The positions of all three commensurability peaks remains quite consistent with the predicted values (grey vertical lines) over a wide range of antidot diameters.

%\subsection{Semiclassical and quantum regimes}
Varying the Fermi wavelength of graphene by gating allows both the classical ($\lambda_F \ll l$, where $l$ is a typical system length scale) and quantum ($\lambda_F \sim l$) regimes to be accessed with the same sample.
In Fig. \ref{fig:efdep}a-d, we examine $R_{xx}$ for the $d\approx10$nm antidot system, with and without geometric disorder, as $\lambda_F$ increases towards the quantum regime.
Red curves in these plots show the cases without geometric disorder -- averaged and individual disordered cases are shown by solid black and light grey curves respectively.
Figure \ref{fig:efdep}a, the case considered throughout this work, shows very little deviation between the curves, suggesting that the peak features are extremely robust against disorder.
As $\lambda_F$ is increased it becomes similar to both $d$ ($\sim 0.4D$) and the antidot neck width ($\sim 0.6D$) and so quantum interference plays a more prominent role.
This is reflected in the increased sample-to-sample fluctuations in disordered cases and a growing discrepancy between pristine and disordered results.
Figure \ref{fig:efdep}e shows local current flow at $B=B_0$ for the largest $\lambda_F$ considered.
`Skipping orbit'-like flows, seen previously for smaller $\lambda_F$, are entirely absent, replaced by a complex pattern of vortices arising from quantum interference effects.

\section{Conclusion}
\label{sec:conc}
We have demonstrated the emergence of magnetic commensurability effects in graphene antidot lattices from purely quantum-mechanical multi-probe transport simulations.
Peak positions and relative magnitudes are in excellent agreement with recent experimental results on the same system.
By examining the dependence of peak positions on cyclotron radius and aspect ratio, and by mapping the local current behaviour at important values, we are able to explain some subtle details of the mechanisms behind commensurability peaks in the longitudinal resistance.
The two highest field commensurability peaks can be understood in terms of scattering between localised states formed around individual antidots in the presence of a magnetic field.
This mechanism is enhanced when these states intersect with nearest, or next-nearest, neighbouring antidots.
Higher order peaks are associated with migration between quasi-pinned orbits around groups of multiple antidots, with scattering from nearby antidots again playing a key role.
We also examined the evolution of the magnetoresistance as different length scales in the system are varied.
The appearance of the second commensurability peak is sensitive to the antidot lattice aspect ratio.
Finally, we demonstrated how the transition between classical and quantum regimes can be tuned by gating.
The onset of the quantum regime at Fermi wavelengths approaching system length scales was associated with the suppression of commensurability features and a marked increase in the effect of geometric disorders.
We believe that our results can act as a guide to interpreting commensurability features in future experiments and, in particular, the role of quasi-bound states in antidot systems.
Our approach can also be easily adapted to consider other perturbation types (e.g. strain\cite{PhysRevB.81.085402, PhysRevLett.117.276801} or gating\cite{Pedersen2012}) or different 2D materials, including bilayer graphene\cite{PhysRevB.85.245426, petersen2015bandgap}.

\section*{Acknowledgments}
We thank A. Sandner and J. Eroms for supplying the experimental data. The authors gratefully acknowledge the financial support from the Center for Nanostructured Graphene (Project No. DNRF103) sponsored by the Danish National Research Foundation and from the \mbox{QUSCOPE} project sponsored by the Villum Foundation. S.R.P. acknowledges funding from the European Union's Horizon 2020 research and innovation programme under the Marie Sk\l{}odowska-Curie grant agreement No 665919,the Severo Ochoa Program (MINECO, Grant No. SEV-2013-0295) and the CERCA Programme / Generalitat de Catalunya.

\appendix
\section{Peierls' phase approach and choice of gauge}
\label{appendix-phases}
The effect of a magnetic field on the behaviour of electrons described within a tight-binding model is included using the Peierls' phase approach.
This involves placing a field-dependent phase factor into the  tight-binding hopping parameters, so that
\begin{equation}
t_{ij} (\mathbf{B}) \rightarrow t_{ij} (0) e ^{\frac{2 \pi i e}{h} \Theta_{ij}},
\label{peierl-hoppings}
\end{equation}
where
\begin{equation}
\Theta_{ij}   = \int_{\mathbf{r}_i}^{\mathbf{r}_j} \mathbf{A}(\mathbf{r}^\prime) \cdot \mathrm{d}
\mathbf{r}^\prime  = \int_0^1 (\mathbf{r}_j - \mathbf{r}_i) \cdot \mathbf{A}(\mathbf{r}_i + \lambda (\mathbf{r}_j - \mathbf{r}_i)) \mathrm{d}\lambda \,.
\end{equation}
\label{peierls-phase}
Note that since the phase factor depends on the vector potential $\mathbf{A}$, where $ \mathbf{B} = \nabla \times \mathbf{A}$, we have freedom with regard to which particular gauge to choose.
For fields perpendicular to the two-dimensional plane ($\mathbf{B} = B\hat{z}$) we usually pick the Landau gauge
\begin{equation}
\mathbf{A}_0 = - B y \hat{x} \,,
\label{starting-gauge}
\end{equation}
which gives
\begin{equation}
\Theta_{ij}  = - B \,  \frac{y_i + y_j}{2}  (x_j-x_i) 
\equiv - B \,  \bar{y} \, \Delta x.
\end{equation}
This gauge proves useful for dealing with leads in the $x$-direction, as the phase term only depends on the difference in $x$ coordinates and not their absolute value -- thus periodic cells can be used within recursive Green's functions methods to calculate the self-energies for leads with this orientation.
Similarly, for leads aligned with the $y$-axis, we can use the gauge $
\mathbf{A}_\perp =  B x \hat{y}$, 
for which
$
\Theta_{ij} =  B \, \bar{x} \, \Delta y 
$,
allowing periodic cells in the $y$-direction and a phase which only depends on the absolute value of the $x$-coordinate.

\begin{figure}
	\centering
	\includegraphics[width =0.49\textwidth]{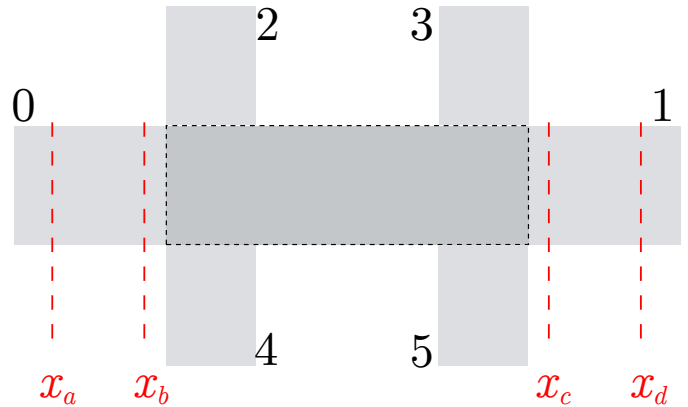}
	\caption{System schematic showing example gauge switching points.}
	\label{fig-schematic-hall}
\end{figure}

For our Hall bar setup, shown schematically in Fig. \ref{fig-schematic-hall}, containing leads with both $x$ ($0,1$) and $y$ ($2,3,4,5$) orientations, we follow the approach outlined by Baranger and Stone\cite{barangerstone}, which allows us to avail of periodicity in every lead of the system.
We add a gauge transformation to the initial gauge in Eq. (\ref{starting-gauge}) so that
\begin{equation}
\mathbf{A} = \mathbf{A}_0 + \mathbf{\nabla} \tilde{f}(x,y) \,.
\end{equation}
It is clear that the magnetic field is gauge invariant under a transformation of this kind.
Ideally we need a $\tilde{f}(x,y)$ which gives $\mathbf{A} = \mathbf{A}_0$ in leads 0, 1 and $\mathbf{A} = \mathbf{A}_\perp$ in leads 2--5.
We define the function
\begin{equation}
f(x, y) \equiv B x y
\end{equation}
and note that
$
\nabla f = B y \hat{x} + B x \hat{y}
$
and
$
\mathbf{A} + \nabla f = \mathbf{A}_\perp \,,
$
allowing us to switch between the two gauges of interest.

We now introduce a weight function $W(x)$ that allows us to smoothly turn on and off the gauge transformation and define
\begin{equation}
\tilde{f}(x,y) = W (x) f (x, y) \,.
\end{equation}
$W$ should take the value $0.0$ in leads $0, 1$ and value $1.0$ in the other leads.
Outside this regions it should vary smoothly between these values with both $W(x)$ and $W^\prime (x) = \frac{\mathrm{d} W}{\mathrm{d} x}$ continuous.
This gives
\begin{equation}
\mathbf{A} = \Big\{ \big( W(x) - 1 \big) \, B y + W^\prime(x) \, B x y \Big\} \hat{x} + W(x) B x \hat{y} \,,
\end{equation}
which we can easily confirm gives the required vector potential when $W(x)=0$ or $1$ and $W^\prime(x) =0$, and further that $\nabla \times \mathbf{A} = B \hat{z}$ at all points.

For our Hall bar system, we include short buffer regions ($x_a < x < x_b$ and $x_c < x < x_d$ in Fig. \ref{fig-schematic-hall}) between the horizontal leads and the central device region, where we switch between the two gauges. We use a smooth step function
\begin{equation}
S_s (z) = 3 z^2 - 2 z,
\end{equation}
where $S_s(0) = 0$, $S_s(1) = 1$, $S_s^\prime(0) = 0$, $S_s^\prime(1) = 0$ to define the weight function in the different regions of our system:
\begin{equation}
W(x) =
\left\{ \begin{array}{@{\kern2.5pt}lL}
0 & if $x \le x_a$.\\
S_s (\frac{x-x_a}{x_b-x_a}) & if $x_a < x < x_b$,\\
1 & if $x_b \le x \le x_c$ \\
1 - S_s (\frac{x-x_c}{x_d-x_c}) &  if $x_c < x < x_d$ \\
0 & if $x \ge x_d$ \,
\end{array}\right.
\end{equation}
%\begin{equation}
%W(x) = \begin{cases}
%0 & x \le x_a \\
%S_s (\frac{x-x_a}{x_b-x_a}) & x_a < x < x_b \\
%1 & x_b \le x \le x_c \\
%1 - S_s (\frac{x-x_c}{x_d-x_c}) &  x_c < x < x_d \\
%0 & x \ge x_d \,.
%\end{cases}
%\end{equation}
With this definition, the phase factors entering in Eq. (\ref{peierl-hoppings}) can be written explicitly as

\begin{widetext}
	\begin{equation*}
		\Theta_{ij} =
		\left\{ \begin{array}{@{\kern2.5pt}lL}
			- B \,  \bar{y} \, \Delta x & if $x \le x_a$.\\
			\begin{array}{l l}
				\frac{B}{2 L_{ab}^3}  &\Bigg[ 2 \Delta y x_i (x_i - x_a)^2 (3L_{ab} - 2x_i + 2x_a)\\
				& - \Delta x \Delta y \Big(L_{ab}^3 + 4 (x_i - x_a)^2 (4x_i - x_a)  - 6 L_{ab} (3x_i^2 - 4x_i x_a + x_a^2\Big) \\
				& - 2 \Delta x y_i \Big(L_{ab} - x_i + x_a \Big) \Big( L_{ab}^2 + L_{ab}(x_i - x_a) - 2(4x_i^2 - 5x_i x_a + x_a^2 )\Big) \\
				& - 4 (\Delta x)^4 (y_i + \Delta y)  + 2 (\Delta x)^3 (3L_{ab} - 8x_i + 6 x_a ) (y_i + \Delta y) \\
				& 	 + 6 (\Delta x)^2 \Big( 3L_{ab}x_i - 4x_i^2 - 2L_{ab}x_a + 6x_i x_a - 2x_a^2\Big)( y_i + \Delta y) \Bigg]
			\end{array} & if $x_a < x < x_b$,\\
			B \, \bar{x} \, \Delta y  & if $x_b \le x \le x_c$ \\
			\begin{array}{l l}
				\frac{B}{2 L_{cd}^3}  & \Bigg[ 2 \Delta y x_i (L_{cd} - x_i + x_c)^2 (L_{cd} + 2x_i - 2x_c)\\
				& + \Delta x \Delta y \Big(L_{cd}^3 + 4 (x_i - x_c)^2 (4x_i - x_c)  - 6 L_{cd} (3x_i^2 - 4x_i x_c + x_c^2\Big) \\
				& + 2 \Delta x y_i \Big(x_i - x_c \Big)   \Big( -9L_{cd}x_i + 8x_i^2 + 3L_{cd} x_c - 10 x_i x_c + 2 x_c^2 \Big) \\
				& + 4 (\Delta x)^4 (y_i + \Delta y)  - 2 (\Delta x)^3 (3L_{cd} - 8x_i + 6 x_c ) (y_i + \Delta y) \\
				& - 6 (\Delta x)^2 \Big( 3L_{cd}x_i - 4x_i^2 - 2L_{cd}x_c  + 6x_i x_c - 2x_c^2\Big)( y_i + \Delta y) \Bigg]
			\end{array} &  if $x_c < x < x_d$ \\
			- B \,  \bar{y} \, \Delta x  & if $x \ge x_d$ \,
		\end{array}\right.
	\end{equation*}
\end{widetext}
The validity of this choice was confirmed by reproducing the quantum Hall effect result for pristine graphene and checking the independence of local current flow on the exact positions of the gauge switching points.

\section{Current maps}
\label{appendix-currents}
The current crossing a single bond between sites $i$ and $j$ is given in its most general form by\cite{caroli1971direct, cresti-currents, nikolic-currents}
\begin{equation}
I_{ij} = \frac{2 e}{h} \int \frac{\mathrm{d} E}{2 \pi} \left( t_{ij} G^{<}_{ji} - t_{ji} G^{<}_{ij}\right),
\end{equation}
where the lesser Green's function can be expressed in term's of the retarded and advanced Green's functions as
\begin{equation}
G^{<}(E) = G^R (E) \Sigma^< (E) G^A (E) \,,
\end{equation}
and
$
\Sigma^< (E) = i \sum_p \Gamma_p (E) f_p (E)
$, 
where $\Gamma_p (E) = i (\Sigma_p (E)- \Sigma^\dagger_p(E))$ is the broadening and $f_p (E) = f (E-eV_p)$ is the Fermi function associated with each lead $p$.

With
\begin{equation}
G^<(E) = i \sum_p f_p (E) G^R (E) \Gamma^p (E) G^A (E)
\label{glesser}
\end{equation}
we can write (omitting $E$ dependence for conciseness)

\begin{equation*}
	t_{ij} G^{<}_{ji} - t_{ji} G^{<}_{ij} = - 2 \, \mathrm{Im} \left[ \sum_p f_p \, t_{ij} \left( G^R \Gamma^p G^A \right)_{ji} \right] \,,
\end{equation*}
where we used $\Gamma^{p} = \Gamma^{p\dagger}$ and $G^A = G^{R\dagger}$.

For the two probe case, with $p=\{L,R\}$ (and assuming $\mu_L > \mu_R$)
%\begin{equation}
\begin{multline}
	\sum_p f_p \,  G^R \Gamma^p G^A  = f_L \,G^R \Gamma^L G^A + f_R \,G^R \Gamma^R G^A 
	\\= i f_R \, (G^R - G^A) + (f_L - f_R) G^R \Gamma^L G^A
	\label{eqn:glessdecomp}
\end{multline}

%\end{equation}
which allows us to split the current into two components:
\begin{itemize}
	\item The \emph{persistent}, or equilibium, current is given by
	\begin{equation}
	I^{(pers.)}_{ij} = \frac{4 e}{h} \int \frac{\mathrm{d} E}{2 \pi} (f_R) \, \mathrm{Re} \left(t_{ij} (G^R - G^A)_{ji}\right) \,.
	\end{equation}
	We note that this component is independent of the chemical potential between leads, and is present even in the absence of a bias. It does not contribute to the total current through the device. Furthermore, the integration involves energies in the Fermi sea below the bias window and it is identically zero in the case of time reversal symmetry.
	\item The \emph{non-equilibrium}, or total current-carrying, component is given by
	\begin{equation}
	I^{(n-eq.)}_{ij} = \frac{-4 e}{h} \int \frac{\mathrm{d} E}{2 \pi} (f_L - f_R) \mathrm{Im} \left( t_{ij} (G^R \Gamma^L G^A)_{ji} \right) \,.
	\end{equation}
	It is carried by electrons at the Fermi surface, and vanishes in the absence of a bias. It is this current that is relevant in this work, and the bond currents plotted in this paper neglect the persistent currents which do not contribute to the overall transport measurements.
\end{itemize}

To generalise to the case of $N$ probes, we split $G^<$ in a similar fashion to Eq. \ref{eqn:glessdecomp} and write
\begin{equation}
\sum_p f_p \,  G^R \Gamma^p G^A  =  i f_m \, (G^R - G^A) + \sum_{p\ne m} (f_p - f_m ) G^R \Gamma^p G^A,
\end{equation}
where $m$ is chosen such that $\mu_m = \mathrm{min} (\mu_{i})$ for $i \in \{0, .., N-1\})$.
The total non-equilibrium bond-current in a multiterminal system is then given by
\begin{equation}
I^{(n-eq.)}_{ij} = \frac{-4 e}{h} \int \frac{\mathrm{d} E}{2 \pi} \sum_{p\ne m} (f_p - f_m) \mathrm{Im} \left( t_{ij} (G^R \Gamma^p G^A)_{ji} \right) \,.
\label{eqn:final_bond_currents}
\end{equation}
In this work we take the low temperature limit so that the Fermi functions become step functions.
Furthermore, we assume that an infinitesimal bias is applied, so that all the quantities in Eq. \ref{eqn:final_bond_currents} are evaluated at the Fermi energy.
Up to a constant, the bond currents are then given by
\begin{equation}
I^{(n-eq.)}_{ij} \sim -\sum_{p\ne m} (V_p - V_m) \mathrm{Im} \, \left( t_{ij} (G^R \Gamma^p G^A)_{ji} \right) \,.
\label{eqn:final_bond_currents2}
\end{equation}

\begin{figure}
	\centering	
	\includegraphics[width =0.5\textwidth]{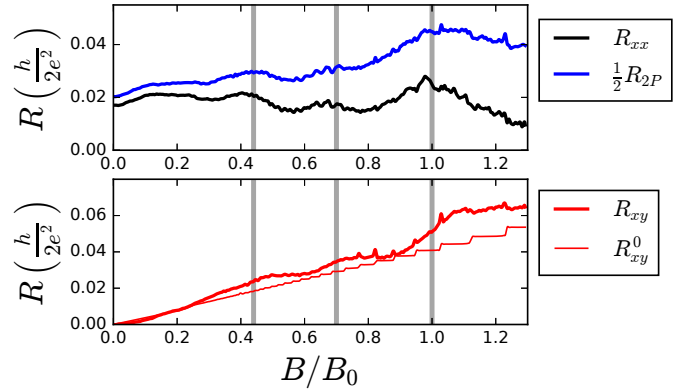}
	\caption{(Top) Comparison of $R_{xx}$ measurement in Fig. 2a from the Hall bar setup (black) with a two-point measurement in a ribbon geometry(blue).
		(Bottom) Corresponding $R_{xy}$ for the antidot system (thick red curve) and for a pristine hall bar without antidot patterning (thin red curve).
	}
	\label{fig:ribbon}
\end{figure}

\section{Longitudinal, Hall and two-point resistances}
\label{appendix-rxy}

In Fig. \ref{fig:ribbon} we compare the longitudinal ($R_{xx}$) resistance from Fig. 2a in the main paper with a two-probe resistance measurement taken for an identical device in a nanoribbon geometry (i.e. without the top and bottom pairs of probes).
We note that although signatures of the commensurability peaks are present in a two-probe measurement, they are significantly less distinct than in the Hall bar device.
The Hall resistance ($R_{xy}$) for the antidot Hall bar device (thick red curve) shows a distinct set of step-like features emerging at the fields corresponding to commensurability peaks in $R_{xx}$\cite{weiss1991electron, Ando2000}.
These steps are not quantised or related to the underlying quantum Hall edge states.
This is clearly seen by comparing the $R_{xy}$ for GALs with that for a pristine nanoribbon (light red curve), where the quantum Hall plateaux are clearly visible and scale very differently in height and width to the new antidot features.

The two-point measurement $R_{2P}$ mixes the peak structure of $R_{xx}$ with the step features of $R_{xy}$.
Furthermore, the Hall bar setup helps to filter out contact resistance effects, which can also obscure commensurability features.
Nonetheless, commensurability features, and particularly the $C_1$ peak, may be visible in two-probe resistances measurements of antidot lattices.

\end{document}